\documentclass[journal,11pt,draftclsnofoot,onecolumn]{IEEEtran}

\pdfoutput=1
\newcommand{\href}[1]{}

\usepackage{subfigure}

\usepackage[colorlinks=true,pdfstartview=FitV,linkcolor=black,citecolor=black,urlcolor=blue,plainpages=false]{hyperref}

\usepackage{cite}
\usepackage[pdftex]{graphicx}

\graphicspath{./figs}

\DeclareGraphicsExtensions{.pdf,.jpeg,.png}
 
\usepackage[cmex10]{amsmath}

\usepackage{array}
\usepackage{mdwmath}
\usepackage{mdwtab}

\usepackage{fixltx2e}

\begin{document}
%\title{Design, Implementation and Characterization of Cooperative Communications Systems}
\title{Design, Implementation and Characterization of a Cooperative Communications System}

\author{Patrick~Murphy and Ashutosh~Sabharwal%
\thanks{The authors are with the Department of Electrical and Computer Engineering at Rice University, Houston, TX, USA, email: \{murphpo,ashu\}@rice.edu. This work was partially funded by NSF grants CNS-0551692, CNS-0619767, CNS-0923479
and CNS-1012921.}
}

\maketitle

\begin{abstract}
Cooperative communications is a class of techniques which seek to improve reliability and throughput in wireless systems by pooling the resources of distributed nodes. While cooperation can occur at different network layers and time scales, physical layer cooperation at symbol time scales offers the largest benefit in combating losses due to fading. However, symbol level cooperation poses significant implementation challenges, especially in synchronizing the behaviors and carrier frequencies of distributed nodes.

We present the implementation and characterization of a complete, real-time cooperative physical layer transceiver built on the Rice Wireless Open-Access Research Platform (WARP). In our implementation autonomous nodes employ physical layer cooperation without a central synchronization source, and are capable of selecting between non-cooperative and cooperative communication per packet. Cooperative transmissions use a distributed Alamouti space-time block code and employ either amplify-and-forward or decode-and-forward relaying.

We also present experimental results of our transceiver's real-time performance under a variety of topologies and propagation conditions. Our results clearly demonstrate significant performance gains (more than 40$\times$ improvement in PER in some topologies) provided by physical layer cooperation, even when subject to the constraints of a real-time implementation. Finally, we present methodologies to isolate and understand the sources of performance bottlenecks in our design.

As with all our work on WARP, our transceiver design and experimental framework are available via the open-source WARP repository for use by other wireless researchers.
\end{abstract}

\section{Introduction}
\label{sec:intro}
Cooperative communications is a mechanism for pooling the resources of distributed nodes to improve the overall performance of a wireless network. Applications of this general idea have been widely studied in the literature, with some of the most prominent~\cite{sendonaris_part1, sendonaris_part2, laneman_coopdiv, laneman2003distributed} having already garnered many thousands of citations. The surveys in~\cite{kramer2006cc,liu2009cooperative} provide excellent overviews of the field from a theory-centric perspective. While cooperative communications has a rich theoretical history in the literature, efforts to actually implement cooperative systems have been much more limited, and thus the issues related to its deployment are still not well understood.

A handful of papers have been published in recent years describing cooperative implementations~\cite{coopImpSurvey_laneman,laneman_coopFramework,berger2005experimental,rice_eurasip,poly_coopImpl_commMag}. Each one, however, falls short of realizing the complete, real-time cooperative transceiver we present in this paper. For example, two implementations are presented in~\cite{poly_coopImpl_commMag}. In the first implementation, the authors focus on cooperation at the MAC layer. This approach is constrained by using standards compliant wireless interfaces whose physical and link layer behaviors cannot be modified in any substantial way. The second implementation uses a software defined radio platform which allows custom physical layer designs. However the software implementation of the physical layer does not operate in real-time, significantly constraining the time scales, achievable data rates and channel conditions which can be evaluated. Finally, in~\cite{laneman_coopFramework} the authors present the performance of a decode-and-forward system built using GNU Radio. They clearly demonstrate a BER improvement using DF, but their transceiver design allows only a single transmission per time slot due to the challenges of synchronizing multiple transmitting nodes. Our design overcomes all these shortcomings, realizing a \emph{complete}, real-time, wideband cooperative transceiver.\footnote{This paper presents a condensed version of the first author's Ph.D. dissertation~\cite{murphyPHD}.}

We describe three key contributions in this paper:

\begin{enumerate}
\item{The design and implementation of a complete cooperative physical layer transceiver. Our design integrates signal processing pipelines, control systems and hardware interfaces into a single FPGA design, realizing a real-time,  cooperative transceiver which operates at bandwidths and time scales comparable to modern wireless networking devices. Further, the transceiver is ready to be integrated with higher layer protocol implementations to study the benefits and implications of employing cooperation in real wireless networks.}

\item{Solutions to two key challenges in realizing real-time physical layer cooperation among distributed nodes, namely the mitigation of multiple carrier frequency offsets during cooperative transmissions and synchronizing the distributed transmissions of cooperating nodes. Our solutions to both are low complexity and are shown to work reliably under a wide range of channel and topological conditions.}

\item{An extensive set of experimental results detailing the performance of our cooperative design. Our experiments test a variety of node topologies and propagation conditions, each designed to model realistic scenarios for modern wireless networking devices. Our results clearly demonstrate substantial performance gains when using cooperation, even when nodes are subject to all the constraints of a complete, real-time design. We also isolate and explain the underlying causes of two error floors observed at high SNR in our experiments.}
\end{enumerate}

The rest of this paper is organized as follows. Section~\ref{sec:design} presents the design of our cooperative transceiver, including discussions of key challenges and our solutions to them. Section~\ref{sec:exp_design} discusses our experimental methodologies. Sections~\ref{sec:isoTri_results}-\ref{sec:linear_results} present end-to-end performance measurements and analysis of the results for two network topologies. Finally, Section~\ref{sec:conclusion} offers concluding remarks.

%%%%%%%%%%%%%%%%%%%%%%%%%%%%%%%%%%%%%%%%%%%%%%%%%%%%%%%%%%%%%%%%%%%%%%%%%%%%%%%%%%%%%%%%%%%%%%%%%%%%%%%%%
%%%%%%%%%%%%%%%%%%%%%%%%%%%%%%%%%%%%%%%%%%%%%%%%%%%%%%%%%%%%%%%%%%%%%%%%%%%%%%%%%%%%%%%%%%%%%%%%%%%%%%%%%
\section{System Design}
\label{sec:design}
This section discusses a few key aspects of our transceiver design, focusing on requirements for building a complete real-time PHY and challenges unique to physical layer cooperation. A description of every subsystem in our transceiver design falls outside the scope of this paper; additional details can be found in~\cite{murphyPHD}, and the source model for the complete transceiver is available open-source~\cite{warpURL}.

\subsection{Overview}
\label{sec:designOverview}
We use the Rice Wireless Open-Access Research Platform (WARP) for all implementation and experiments in this work. Our cooperative transceiver is implemented as a custom FPGA core using Xilinx System Generator~\cite{sysgen_url}, designed to run in the FPGA at the heart of the WARP hardware. The transceiver core implements complete OFDM transmit and receive pipelines, each with digital I/Q interfaces at one end and a packet buffer at the other. All signal processing and control subsystems are implemented in the FPGA fabric and run in real-time; the transceiver does not rely on any processing external to the FPGA at each WARP node. The design currently operates with 64 subcarriers in 10~MHz RF bandwidth (scalable to 20~MHz), achieving payload datarates of 6/12/24 Mbps for BPSK/QPSK/16-QAM modulation rates, respectively.

\noindent{\bf Distributed STBC:} We focus on cooperative schemes which utilize simultaneous source and relay transmissions in the same frequency band. These overlapping transmissions are orthogonalized using a distributed version of the Alamouti space-time block code~\cite{Alamouti}, wherein the single-antenna source and relay nodes seek to behave as two transmit antennas in a MIMO transmitter. This design provides the significant benefit of a receiver which does not require a priori knowledge of whether one or two nodes are transmitting any given packet. However, the use of simultaneous transmissions by the source and relay imposes two major challenges: mitigating carrier frequency offsets among the three nodes and synchronizing the transmissions of the source and relay. These challenges, and our solutions to each, are discussed in Sections~\ref{sec:cfo} and~\ref{sec:sync} below.

\noindent{\bf Relaying Modes:} In a cooperative system the behavior of a relay can be broadly classified by how much processing it applies to a received waveform before retransmitting it. At one extreme is amplify-and-forward (AF), in which the relay applies no processing; it simply records the received waveform and re-transmits it. At the other extreme is decode-and-forward (DF), where the relay implements both a physical layer receiver, for decoding the source's transmission, and transmitter, for re-transmitting the decoded payload. Our cooperative transceiver implements both AF and DF relaying and can switch between modes per-packet.\footnote{While not necessary to characterize transceiver performance per-scheme, this feature directly enables future implementation of cooperative MAC protocols which can dynamically switch between cooperative schemes.} For AF, the waveform capture is implemented digitally, recording the unprocessed I/Q samples at the input to the PHY receiver. Many other relaying schemes have been proposed~\cite{kramer2006cc} which employ other kinds of processing at the relay (estimate-and-forward, quantize-and-forward, etc.); we focus on AF and DF for both their tractable implementations and straightforward interoperability with distributed space-time coding.

\noindent{\bf Receiver End States:} An important part of our complete transceiver design is a receiver front-end which enables autonomous operation. This front-end is responsible for detecting energy events which may indicate an incoming transmission, controlling the radio's gain settings (AGC) and establishing sample-level synchronization in the received waveform. Only after these processes are complete can the back-end of the PHY receiver begin processing the waveform in its attempt to decode the header and payload of the incoming packet.

\begin{figure}[h]
\centering
\includegraphics[width=2in]{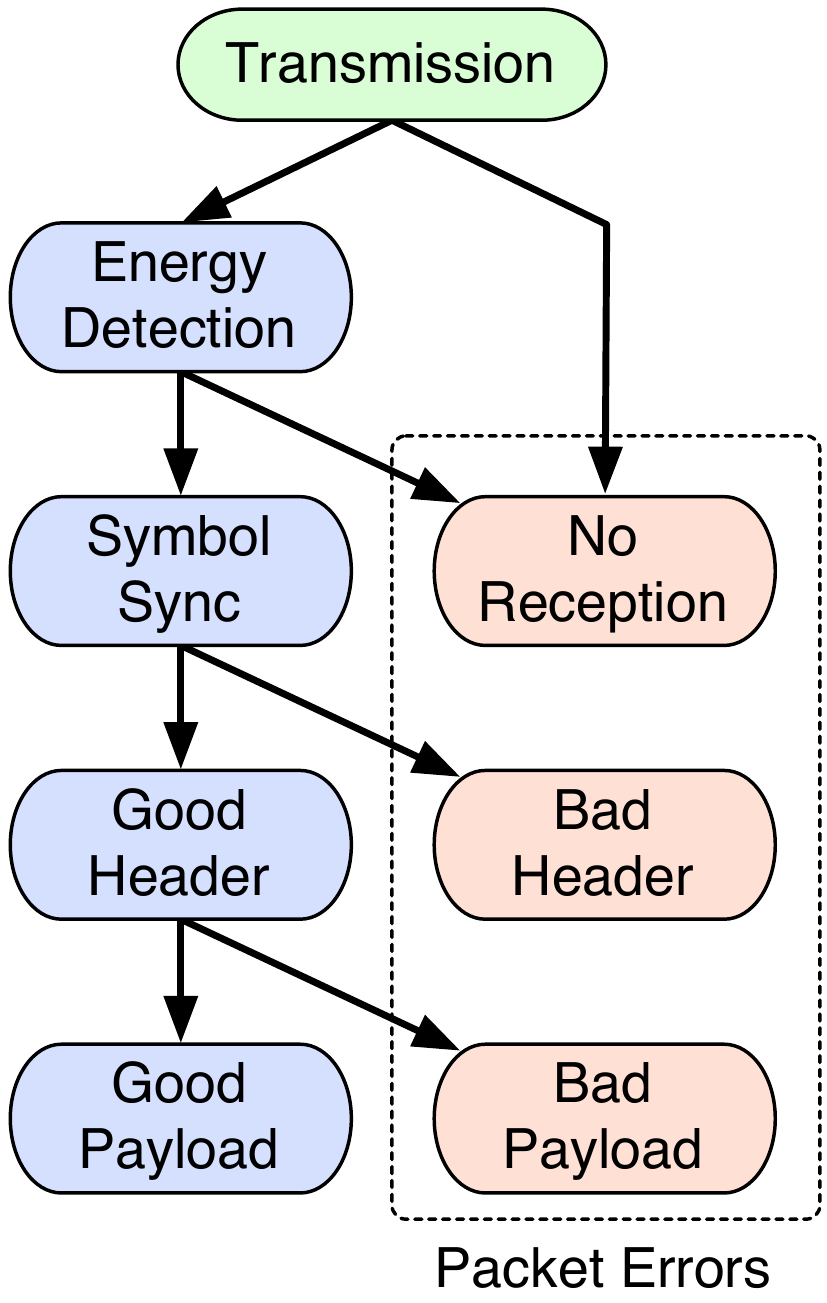}
\caption{State transitions and possible outcomes of a given packet transmission.}
\label{fig:rxOutcomes}
\end{figure}

Each stage of the receiver's processing is critical; a failure at any stage will result in a packet error. Fig.~\ref{fig:rxOutcomes} illustrates the potential outcomes of a given transmission as it progresses through the receiver's various processing stages. This chart starts with a transmission and ends in one of four states. Only the \textsf{Good Payload} state represents a successful reception. The other three end states represent packet errors. The \textsf{No Reception} state represents failures in energy detection or synchronization. When energy detection and synchronization succeed, the receiver attempts to decode the packet header and payload. If bit errors occur in the header, the receiver terminates in the \textsf{Bad Header} state and does not attempt reception of the payload (an invalid header precludes using its rate and length fields to configure the receiver for processing the payload). Finally, the receiver will terminate in the \textsf{Bad Payload} state if payload bit errors are detected, which can only occur if energy detection and header decoding have already succeeded.

It is important to recognize that, from the receiver's perspective, payload bit errors only occur in packets terminating in the \textsf{Bad Payload} state. The observation that packet errors can occur with no bit errors (i.e. when no attempt is made to decode a payload) is key in understanding the performance results presented in Sections~\ref{sec:isoTri_results}-\ref{sec:linear_results}.

\subsection{Carrier Frequency Offset}
\label{sec:cfo}
Carrier frequency offset (CFO) is a common impairment in real wireless systems. It results from variations in frequency across the local oscillators employed by communicating nodes to generate the carrier signals they use for translating signals between baseband and RF. The issues of CFO are well studied in wireless systems. However, the impact of CFO and techniques to mitigate it depend heavily on the specific parameters of a given transceiver and the properties of the hardware on which it is realized. Our OFDM receiver uses an adaptation of a standard technique~\cite{schmidl_ofdmSync} for estimating and correcting CFO in the time domain. This scheme exploits periodicity in each packet's preamble to estimate CFO and significantly reduces the effective CFO before samples are fed into the FFT. However the receiver still experiences phase offsets in the frequency domain due to the residual CFO. We use an adaptation of the scheme proposed in~\cite{phaseTracking_sakata} to track and correct these errors. These systems (time-domain CFO correction and frequency-domain phase correction) are used during receptions of both non-cooperative and cooperative transmissions.

Physical layer cooperation imposes significant additional challenges in mitigating the effects of CFO. Specifically, when two nodes transmit simultaneously, the physical layer design must consider both the offset between the transmitters and each transmitter's offset relative to their common destination (labeled $\Delta{f_{SD}}$, $\Delta{f_{SR}}$ and $\Delta{f_{RD}}$ below). There are two general ways of dealing with these multi-CFO issues. The first burdens the destination's receiver with estimating both $\Delta{f_{SD}}$ and $\Delta{f_{RD}}$ and mitigating their combined effects. A number of proposed schemes take this approach~\cite{huang2009data, li2008carrier, mietzner2005distributed}. All of these schemes require significant increases in the complexity of the receiver implementation. For example the scheme in~\cite{li2008carrier} employs substantial processing in the time domain which requires estimation of the channel impulse response (not generally available in an OFDM receiver) and estimates of both $\Delta{f_{SD}}$ and $\Delta{f_{RD}}$. Further, this scheme adds significant signaling overhead and requires the destination know a priori which combination of source and relay nodes are participating in every transmission.

Our design focuses on an alternative approach, in which the problem of multiple CFOs during cooperative transmissions is mitigated entirely at the relay. This approach seeks to mimic the CFO behavior in a non-distributed 2$\times$1 Alamouti link, where the carrier frequencies of the two transmissions (originating from a single two-antenna transmitter), are identical. In our cooperative system the relay must assure its transmission occurs with zero frequency offset relative to the source so that, from the destination's perspective, there is no frequency offset between the two simultaneous transmissions. Pre-correcting the CFO at the relay provides two key benefits. First, no extra processing is required at the destination; it needs to estimate only one CFO ($\Delta{f_{SD}}$), preserving the very useful feature of a receiver design which does not require knowledge of which combination of source and relay are participating in a given transmission. Second, no extra overhead is required; the relay extracts everything it requires to pre-correct its own transmission from the source's normal transmission. This approach requires the relay know the frequency offset between its own transmit and receive paths. In many devices, including the WARP hardware we use for our experiments, this offset is zero as the RF circuits in both paths use a common reference oscillator.

In an amplify-and-forward relay the frequency pre-correction occurs automatically~\cite{rice_eurasip}; the frequency offset incurred through the radio receiver is inherently reversed when the unmodified received signal is fed back through the radio transmitter. A decode-and-forward relay requires more care to achieve the necessary carrier frequency matching. Our DF relay design handles CFO in two stages. In the first time slot of a cooperative transmission the DF relay estimates $\Delta{f_{SR}}$. In the second time slot the relay multiplies its transmitted waveform by a complex sinusoid at frequency $-\Delta{f_{SR}}$. The sinusoid generation and multiplication are implemented digitally in the final stage of the OFDM transmit pipeline.

\begin{figure}[h]
\centering
\includegraphics[width=3.5in]{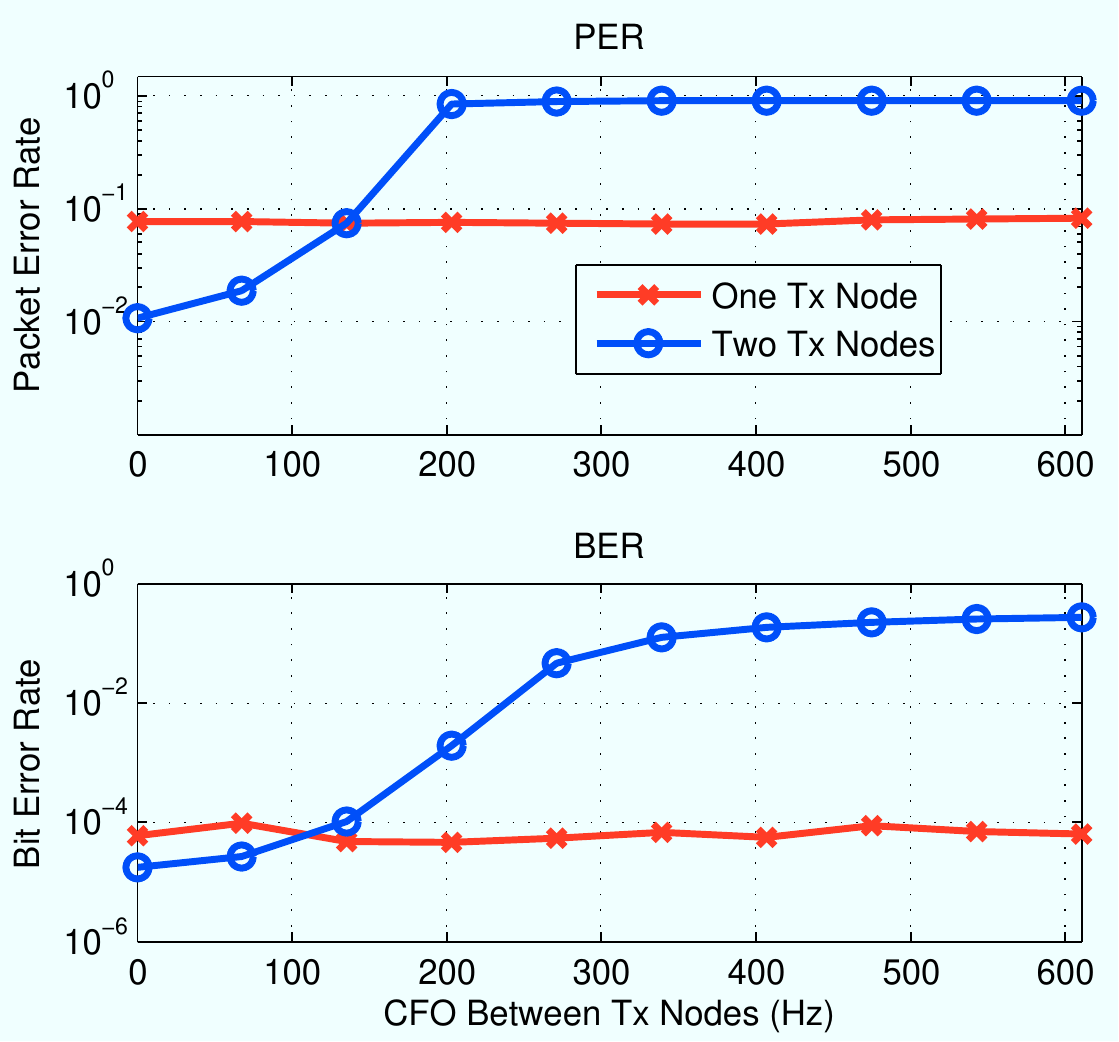}
\caption{Experimental PER/BER performance vs. CFO between transmitting nodes.}
\label{fig:berper_DF_vs_SRcfo_noPrespin}
\end{figure}

The performance of this explicit CFO pre-correction scheme for DF depends on the quality of the relay's $\Delta{f_{SR}}$ estimate. Fig.~\ref{fig:berper_DF_vs_SRcfo_noPrespin} illustrates just how accurate this estimate must be to achieve any reasonable performance. These plots show measurements of end-to-end PER and BER as a function of the frequency offset between the source and relay; in effect, the independent variable here is the relay's error in estimating $\Delta{f_{SR}}$. In this experiment three WARP nodes (S, R and D) are connected via a channel emulator (discussed below), configured for frequency flat fading and high average SNR along all three paths (SR, SD and RD). The ``One Tx Node" curves show performance when the relay is inactive. The ``Two Tx nodes" curves show the performance with decode-and-forward relaying. The OFDM waveform here consists of 64 subcarriers in 10~MHz bandwidth (156~kHz subcarrier spacing), using the same PHY design as in all the experiments discussed below.

In the hardware setup for this experiment the source and relay nodes share a frequency reference, assuring zero offset between their actual carrier frequencies. The relay applies a frequency offset digitally in the final stages of its transmit pipeline. By sweeping the value of this offset we can induce a known difference between the center frequencies of the source and relay transmissions, perfectly mimicking $\Delta{f_{SR}}$ estimation errors in the relay's CFO pre-correction system. The resulting PER/BER measurements clearly demonstrate the challenge of building an effective CFO pre-correction system at the relay. 

\begin{figure}[h]
\centering
\includegraphics[width=3.0in]{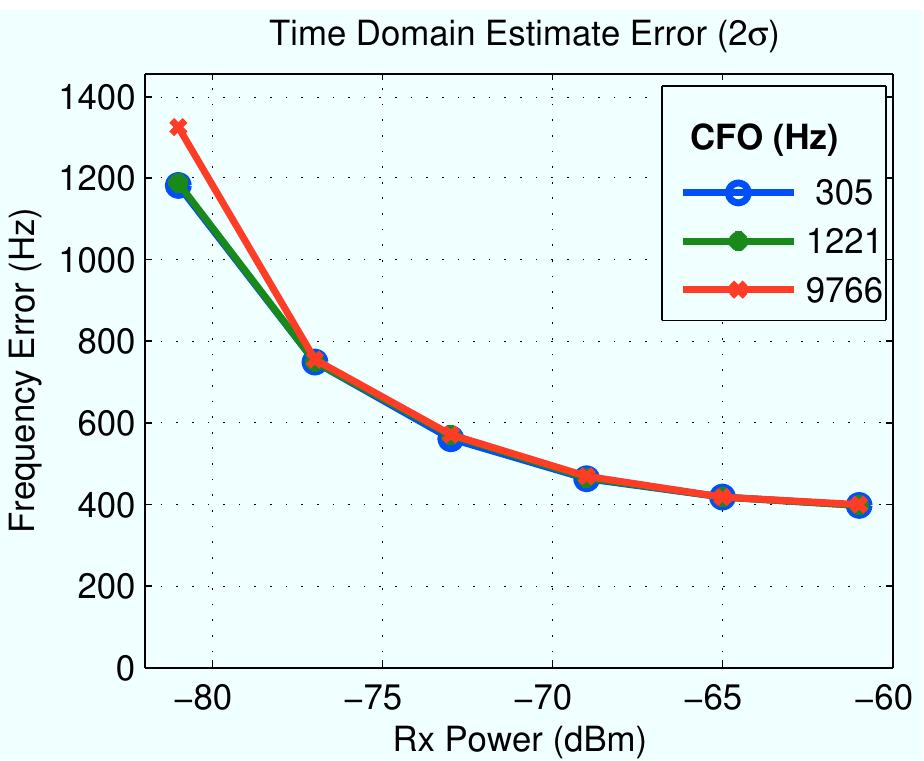}
\caption{Measured performance of the time-domain CFO estimator as a function of SNR.}
\label{fig:cfoEstError_coarse_vs_snr}
\end{figure}

For small values of $\Delta{f_{SR}}$ (i.e. small estimation errors) the relay provides a performance gain over the non-cooperative link. However as $\Delta{f_{SR}}$ increases the performance of the cooperative link drops significantly, degrading to nearly 100\% PER for offsets larger than about 200 Hz. This transition from performance improvement to guaranteed error establishes the very tight tolerance for estimation errors in the CFO pre-correction system at the relay. This estimation error tolerance is so small, in fact, that the existing time-domain CFO estimator (calculated using the preamble, as discussed above) is insufficiently accurate for use here. Fig.~\ref{fig:cfoEstError_coarse_vs_snr} shows the performance of the time-domain estimator as a function of SNR and CFO, plotted as twice the standard deviation (2$\sigma$) of the estimate distributions (the estimate distributions were consistently Gaussian with means at the actual frequency offset). These measurements were taken between two nodes connected by coaxial cables and a variable attenuator (i.e. effectively an AWGN link). When compared to the CFO estimation error tolerance shown in Fig.~\ref{fig:berper_DF_vs_SRcfo_noPrespin}, the time-domain estimator's variance is far too high for use in CFO pre-correction at the relay.

%coop_testing22\v20_cfoCar_LOCAL

\begin{figure}[h]
\centering
\includegraphics[width=3.0in]{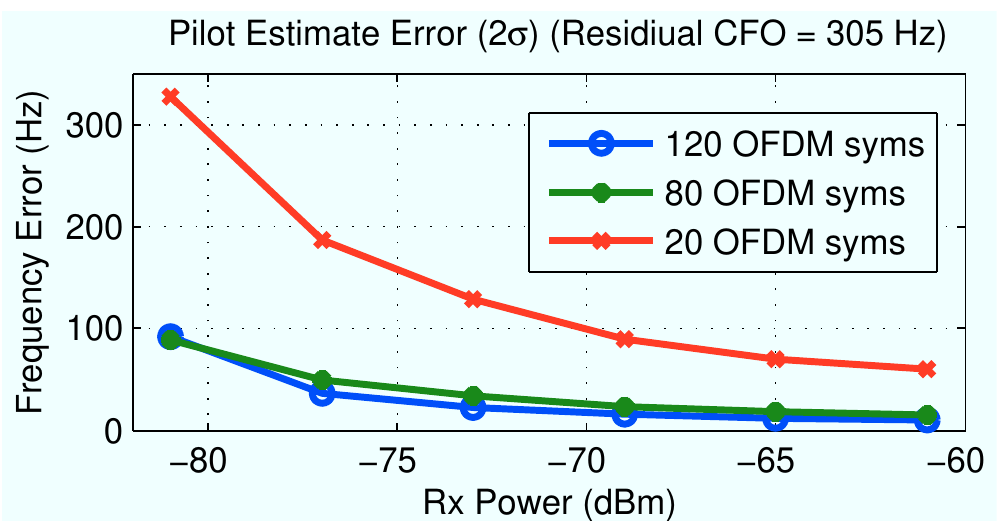}
\caption{Experimental performance of the frequency-domain residual CFO estimator vs. SNR for multiple packet durations.}
\label{fig:cfoest_dist_pilot}
\end{figure}

Our cooperative receiver design includes a second estimator which refines the CFO value for pre-correcting relay transmissions. Unlike the time-domain CFO estimator, which must provide a valid estimate very early in a packet reception, the secondary CFO estimator is able to use the full received waveform to generate its estimate, as its estimate is required only before the next transmission begins. We exploit this relaxed timing requirement by calculating the secondary CFO estimate using the pilot tones embedded in each OFDM symbol. Specifically, we calculate the residual CFO in each packet as the ratio of the phase of the pilot tones in the final OFDM symbol and the packet duration. This simple calculation provides a very accurate CFO estimate, as illustrated in Fig.~\ref{fig:cfoest_dist_pilot}. This plot shows the performance of the pilot-based CFO estimator as $2\sigma$ of the distribution of estimates verses both SNR and packet length (the estimate distributions were consistently Gaussian centered with means at the actual frequency offset). In this experiment the time-domain CFO estimator is disabled, the frequency offset between the transmit and receive nodes is fixed at 305 Hz (modeling a nominal residual CFO) and the nodes are again connected via cables and a variable attenuator.

%coop_testing22\v20_cfoCar_LOCAL

When compared to the CFO estimation tolerance established in Fig.~\ref{fig:berper_DF_vs_SRcfo_noPrespin}, it is clear this estimator should perform well enough for use in pre-correcting CFO at the relay during cooperative transmissions. This expectation is confirmed via the extensive performance evaluations of the DF link, as discussed below.

\subsection{Synchronization}
\label{sec:sync}
Achieving useful simultaneous transmissions by distributed nodes imposes a significant synchronization challenge. Cooperating transmitters must initiate their transmissions nearly simultaneously to best mimic transmissions from two antennas in a MIMO transmitter.\footnote{We intentionally ignore differences in SD and RD propagation times here, recognizing our nodes are always within one sample period of propagation time ($\approx$100 feet at 10 MHz bandwidth) of one another. However, to support future experiments with larger topologies, our PHY design includes a programmable transmit delay with 25 nsec resolution which can be updated per-transmission.} Further, distributed nodes without access to a central synchronization source must align their transmissions using only information received over the same wireless interfaces used for actual payloads.

We address this requirement with a dedicated subsystem in the PHY which can automatically initiate a packet transmission either a fixed period after a previous transmission or in response to a packet reception. Implementing this functionality in the FPGA fabric results in identical latencies for the Tx$\rightarrow$Tx or Rx$\rightarrow$Tx turnaround times at every node. The source uses this subsystem to re-transmit its packet in the second time slot of each cooperative exchange. The relay uses it to initiate its transmission after receiving the source's transmission in the first slot. With deterministic timing at both source and relay, their transmissions are consistently aligned within a single sample period (100 nsec), the best which can be expected given each node uses its own sampling clock.

This auto-transmit subsystem also directly enables cooperative MAC designs. In~\cite{doc_ciss}, for example, we use this feature to synchronize source and relay transmissions when each is independently triggered by reception of a control packet from the destination node. The conditions for initiating a transmission and contents of the transmitted packets are programmable at run-time, preserving both deterministic timing for PHY synchronization and full flexibility in MAC protocol design. More details about the design and application of this system are available in~\cite{murphyPHD,warpOFDM_assilomar,doc_ciss}.

%%%%%%%%%%%%%%%%%%%%%%%%%%%%%%%%%%%%%%%%%%%%%%%%%%%%%%%%%%%%%%%%%%%%%%%%%%%%%%%

\subsection{Design Integration}
The cooperative OFDM transceiver is just part of the overall node design. Additional FPGA cores are required to manage hardware peripheral interfaces (Ethernet, radio transceiver, etc.) and to connect each peripheral to a common bus managed by an embedded PowerPC processor. This processor executes C code consisting of both drivers for each core and the WARPMAC framework for implementation of custom MAC protocols~\cite{warpURL}.

The logic designs, C code and hardware interfaces are all integrated via Xilinx Platform Studio (XPS), which serves as a front-end for both the logic synthesis flow and software compilation flow. The output of an XPS project is a single bitstream used to configure the WARP hardware's FPGA. We use a common bitstream for all three nodes in our experiments. The role of each node (source/relay/destination) is defined by the position of a switch on the FPGA board, effectively determining the node's MAC address at boot. This design flow provides a key benefit, in that any node can assume any role at run-time. The PHY actually supports changing roles per-packet. We use fixed roles in the PHY characterization experiments discussed below, but the design is ready for experiments with cooperation-aware MAC protocols which switch roles on the fly.

%%%%%%%%%%%%%%%%%%%%%%%%%%%%%%%%%%%%%%%%%%%%%%%%%%%%%%%%%%%%%%%%%%%%%%%%%%%%%%%%%%%%%%%%%%%%%%%%%%%%%%%%%
%%%%%%%%%%%%%%%%%%%%%%%%%%%%%%%%%%%%%%%%%%%%%%%%%%%%%%%%%%%%%%%%%%%%%%%%%%%%%%%%%%%%%%%%%%%%%%%%%%%%%%%%%

\section{Experiment Design}
\label{sec:exp_design}
As discussed in Section~\ref{sec:intro}, two of our primary goals are the design of a cooperative physical layer transceiver (discussed in Section~\ref{sec:design}) and a thorough evaluation of the transceiver under a variety of conditions (discussed in Sections~\ref{sec:isoTri_results}-\ref{sec:linear_results}). Connecting these goals is the need for experimental methodologies and experimental parameter selection to measure the performance of our cooperative transceiver implementation.

Addressing these requirements poses significant design challenges. For example, any cooperative experiment requires coordination of three nodes, acting as source, relay and destination. The experiment must account for every source transmission, relay reception, relay transmission and destination reception, and the conditions under which each takes place. Further complicating the experimental requirements is the need to control the wireless propagation environment. We need to test a variety of SNRs and fading conditions and must do so reliably and repeatably. This section discusses our solutions to these challenges.

\subsection{Channel Emulator}
For our experiments we use an Azimuth ACE 400WB wireless channel emulator~\cite{azimuth_400wb}. The emulator interfaces to wireless devices using coaxial cables, mimicking (from the wireless node's perspective) an actual antenna connection. The emulator accepts and generates signals at the same power levels as antennas, allowing wireless devices under test to behave exactly as if they were communicating over the air.

Our experiments use three WARP nodes, each with a single RF interface connected to the channel emulator. We use the emulator to independently control the properties of three channels (SD, SR \& RD). Each channel is configured with fading statistics and an average path loss. The fading statistics are configured via selection of a channel model. 
We use two models from Azimuth library originally developed by TGn Sync~\cite{80211n_chanModels} (one of the groups which merged to form the IEEE 802.11n working group). We use TGn Model A, which implements frequency flat fading and TGn Model B, which implements frequency selective fading with a 15 nsec RMS delay spread. In all experiments the fading velocity is configured at its maximum of 1.2 km/h. The average path loss per channel is the sum of the inherent attenuation through the emulator and a programmable attenuator at each RF output. In our setup the inherent attenuation is 53 dB, and the programmable attenuator value, the independent variable in most of the plots below, is configured between 0 and 36 dB.

\subsection{Topologies}
The relative positions of the source, relay and destination nodes are important factors in determining the performance of a cooperative link. We emulate various network topologies by varying the average path loss along each of the three emulated channels.

\begin{figure}[h!]
\centering
\subfigure[Co-located source/relay topology]{\includegraphics[width=2.5in]{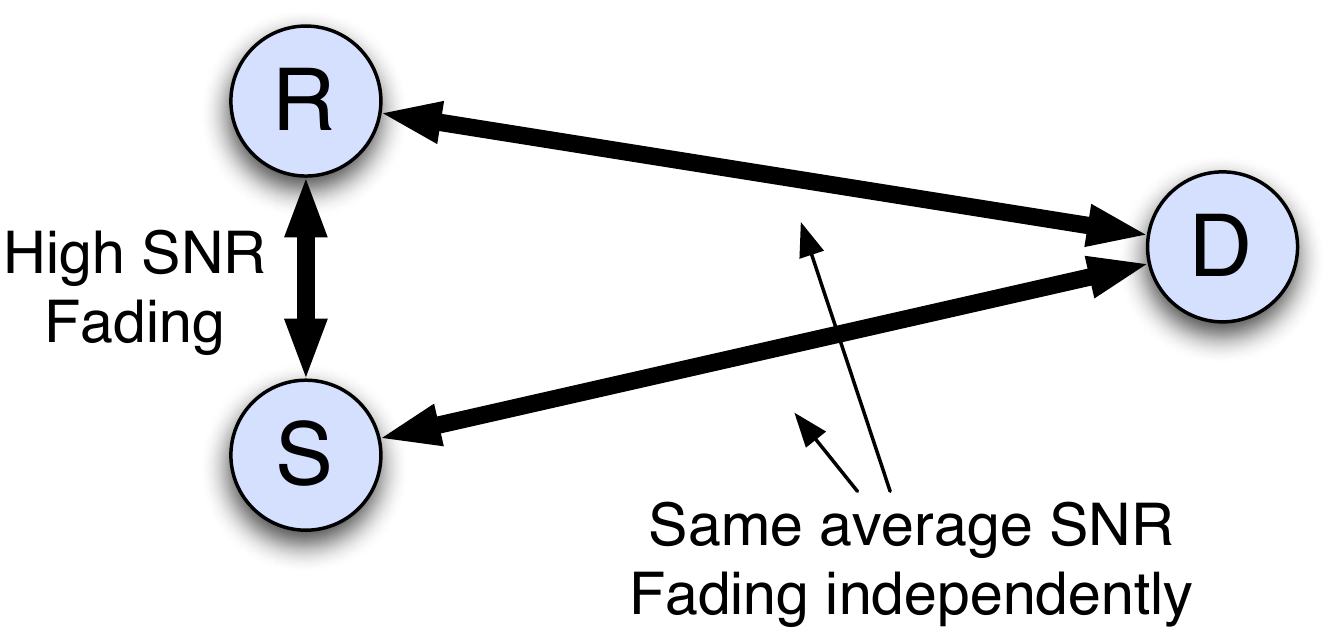}}
\subfigure[Linear topology]{\includegraphics[width=2.9in]{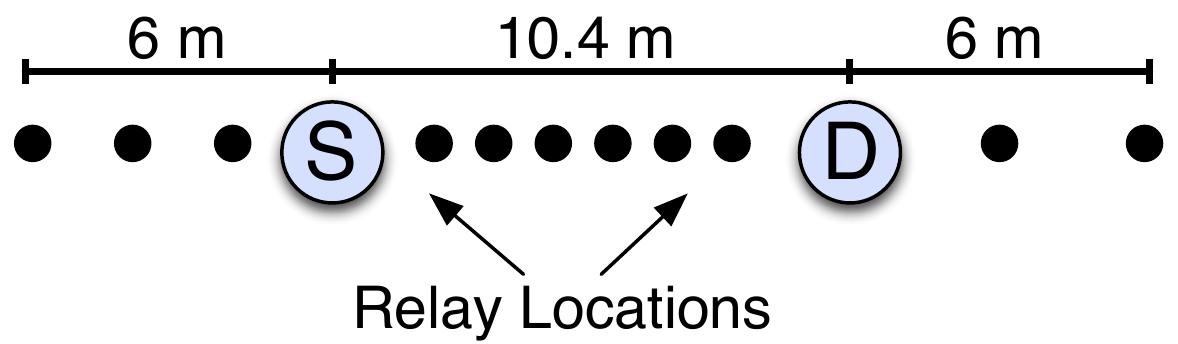}}
\caption{Emulated network topologies for performance measurements.}
\label{fig:topologies}
\end{figure}

We present results from two emulated topologies here (results with additional topologies are available in~\cite{murphyPHD}). The first, illustrated in Fig.~\ref{fig:topologies}(a), models a co-located source and relay node, with the destination node some distance away. All three wireless links in this topology are subject to random fading. The average SR path loss is fixed at the minimum value possible through the channel emulator ($\approx$53~dB). This topology is parameterized by the equal average SD and RD path losses; these are swept over a 36~dB range in our tests. This topology tests an interesting usage case for cooperation. Privacy and incentivizing participation are common concerns with employing cooperation in real networks. These issues become much easier if a single user owns the devices participating in a cooperative transmission. For example, one person's co-located laptop and phone (with compatible wireless interfaces) could cooperate to improve communications with a base station.

The second emulated topology, illustrated in Fig.~\ref{fig:topologies}(b), is parameterized by the position of the relay relative to a fixed source and destination. All three links are subject to random fading. The average SD path loss fixed at 71~dB, modeling a distance of $\approx$10 meters.\footnote{The mapping of path loss to distance requires selection of a path loss exponent. We use 2.1 (modeling indoor propagation) for all distance-path loss mappings in this work. A different exponent would change only the topological interpretation of average path losses, not our measured performance at each.} The average SR and RD path losses are swept together to model various relay positions along the line connecting the source and destination nodes.

\subsection{Metrics}
\label{sec:metrics}
Our experiments are focused on characterizing the performance of the cooperative physical layer transceiver discussed above. Two primary metrics are useful here: packet error rate and bit error rate.

\noindent{\bf Packet Error Rate}: Recall from Section~\ref{sec:designOverview} that in our design every transmission ends in one of four outcomes at the receiver, and that only a \textsf{Good Payload} result represents a successful reception. Thus, we classify a packet error as any transmission which does not end in the \textsf{Good Payload} state at the intended destination. This leads to our definition of packet error rate, $(\mathsf{1-\frac{N_{RxGood}}{N_{Tx}}})$, where $\mathsf{N_{RxGood}}$ is the number of packets received without error (i.e. ended in the \textsf{Good Payload} state) and $\mathsf{N_{Tx}}$ is the number of transmitted packets. 

Recall that the various failure states (\textsf{No Reception}, \textsf{Bad Header} and \textsf{Bad Payload}) all count as packet errors. Our design tracks how many packets end in each state independently. This capability proves very useful in analyzing the overall PER and BER results for a given trial, helping identify the dominant source of errors in various regimes. The discussion of results in Sec~\ref{sec:resultsDiscussion} explores this further.

\noindent{\bf Bit Error Rate}: Bit error rate is a widely used metric for understanding PHY performance. From the perspective of MAC protocols, any bit error is unacceptable; packets decoded with 1 or 5000 bit errors are equally useless for higher network layers. But knowing how many bit errors are responsible for a bad packet is very useful in gauging performance of a PHY and in designing error correcting codes.

Our OFDM receiver interprets header fields to determine the payload length and modulation rate for each reception. If an error is detected in the header itself (the header has its own checksum) the receiver halts and does not attempt to receive the payload, being unsure of the parameters needed to process it. This is the \textsf{Bad Header} outcome illustrated in Fig.~\ref{fig:rxOutcomes}. Thus, our design experiences payload bit errors only for receptions which end in the \textsf{Bad Payload} state.

This receiver design leads to a definition of bit error rate as the ratio $\mathsf{\frac{B_{Error}}{B_{Total}}}$, where $\mathsf{B_{Error}}$ is the number of payload bit errors and $\mathsf{B_{Total}}$ is the total number of payload bits processed. Note that $\mathsf{B_{Total}}$ includes bits from packets which end in both the \textsf{Good Payload} and \textsf{Bad Payload} states. Header bits do not count towards either value. Packets which are not detected (\textsf{No Reception} outcome) or those received with header errors (\textsf{Bad Header} outcome) do not contribute to the BER calculation. As a result, our BER measurements must be considered in tandem with the corresponding packet error rates to fully understand the transceiver performance.

%%%%%%%%%%%%%%%%%%%%%%%%%%%%%%%%%%%%%%%%%%%%%%%%%%%%%%%%%%%%%%%%%%%%%%%
%%%%%%%%%%%%%%%%%%%%%%%%%%%%%%%%%%%%%%%%%%%%%%%%%%%%%%%%%%%%%%%%%%%%%%%

\section{Experiment Results: Co-located Source/Relay}
\label{sec:isoTri_results}

\begin{figure}[h!]
\centering
\subfigure[QPSK payload modulation]{\includegraphics[width=2.9in]{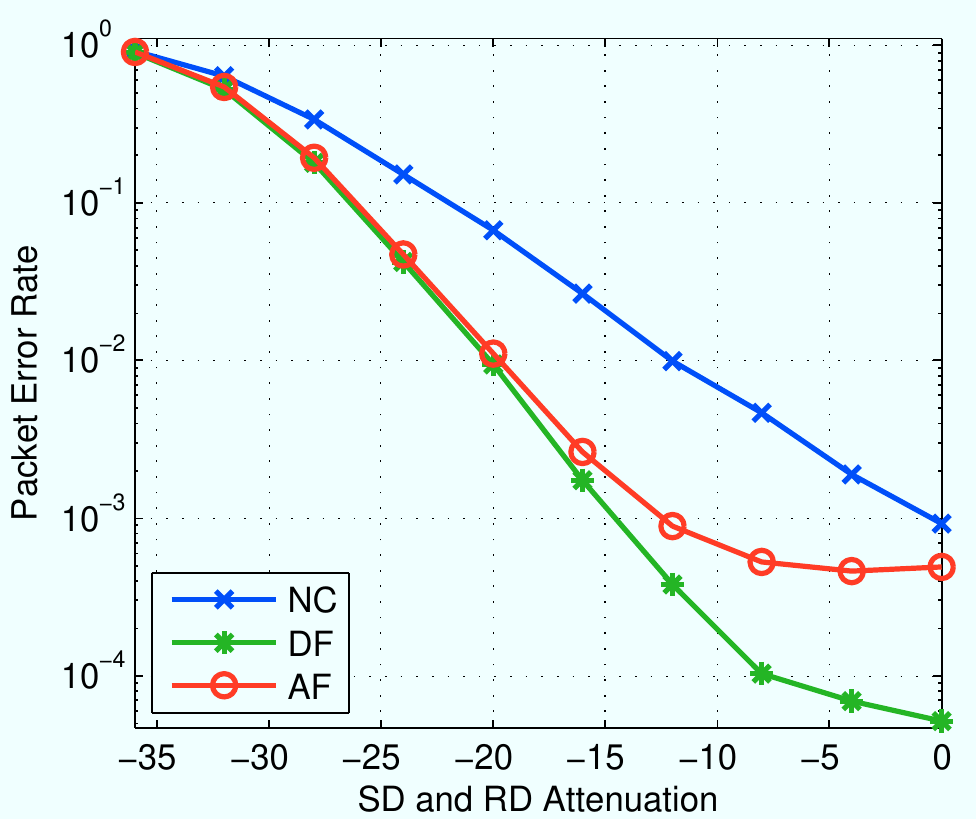}}
\subfigure[16-QAM payload modulation]{\includegraphics[width=2.9in]{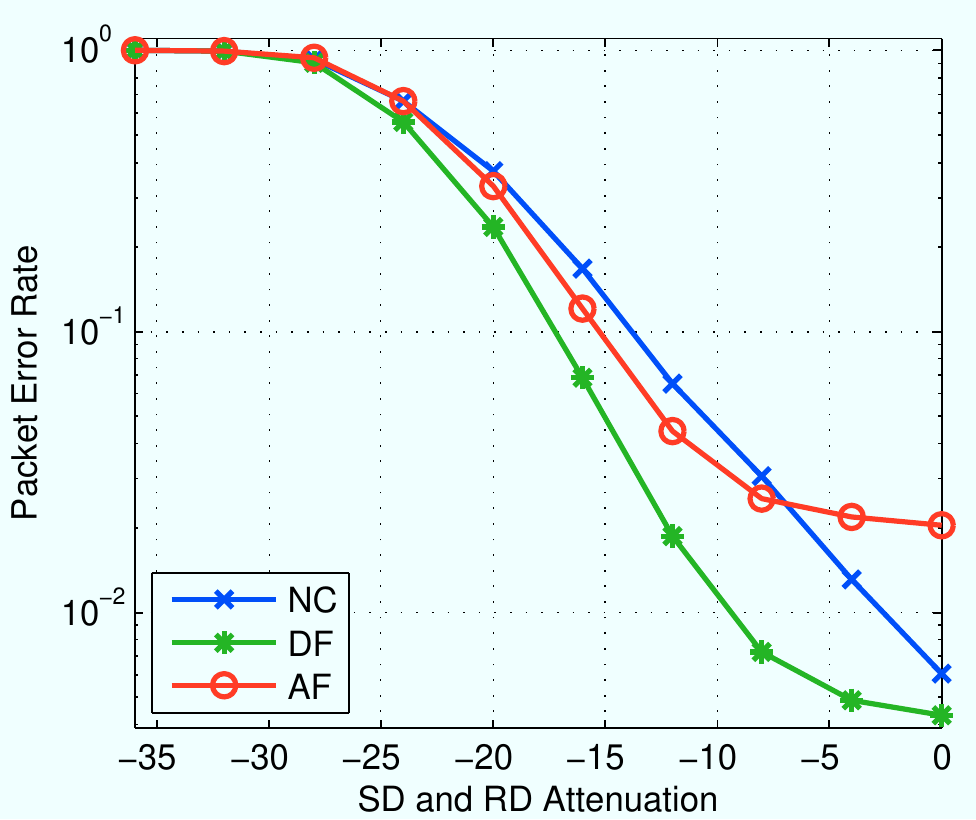}}
\caption{Packet error rates for co-located source/relay topology with 1416 byte payloads in frequency flat fading.}
\label{fig:isoTri_per_flat}
\end{figure}

\begin{figure}[h!]
\centering
\subfigure[QPSK payload modulation]{\includegraphics[width=2.9in]{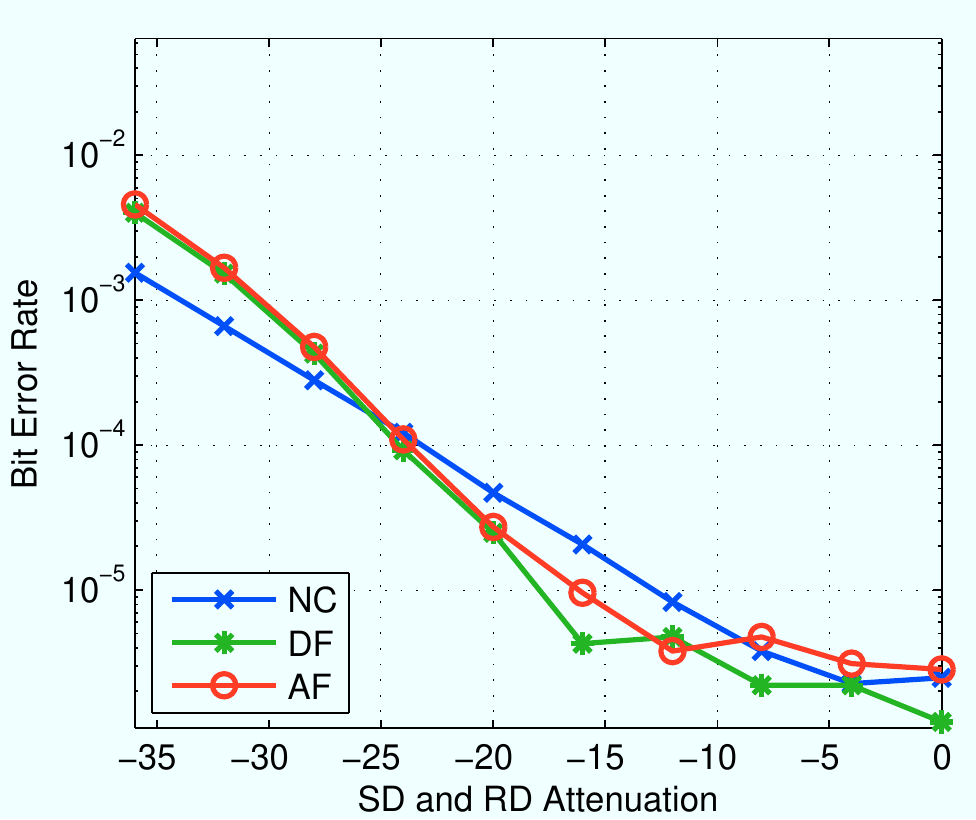}}
\subfigure[16-QAM payload modulation]{\includegraphics[width=2.9in]{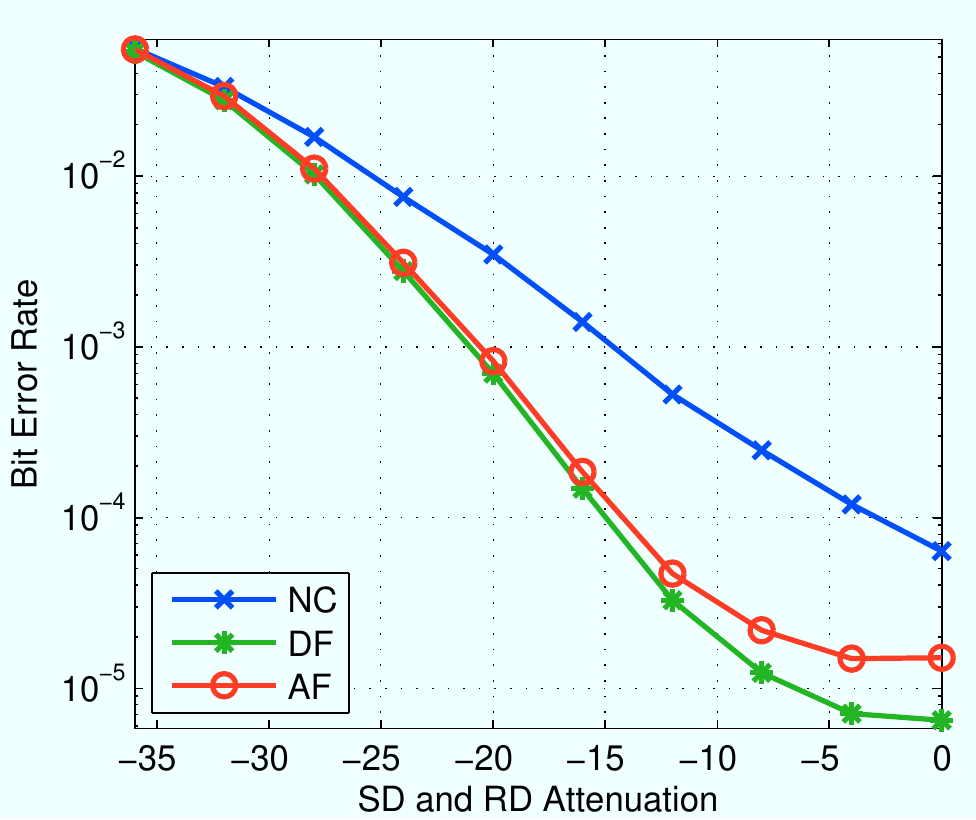}}
\caption{Bit error rates for co-located source/relay topology with 1416 byte payloads in frequency flat fading.}
\label{fig:isoTri_ber_flat}
\end{figure}

This section presents extensive PER and BER measurements for the co-located source/relay topology, gathered during experiments using the transceiver design, node configuration and channel emulator setup described above. The parameters for this experiment are:
\begin{itemize}
\item{{\bf Topology points:} The SR attenuation is always zero, giving an average SR path loss of 53 dB (the inherent path loss through the emulator). The SR and RD attenuations are always equal and are swept from 0 to 36 dB in 4 dB steps (53 to 89 dB total average path loss).}

\item{{\bf Packet formats:} Every transmitted packet consists of a random 1412 byte payload, modulated with either QPSK or 16-QAM. Every packet has a 24 byte header modulated with QPSK. The header and payload are both uncoded.}

\item{{\bf Cooperative schemes:} This experiment tests three schemes: non-cooperative (no relay participation; labeled {\bf NC}), amplify-and-forward relaying (labeled {\bf AF}) and decode-and-forward relaying (labeled {\bf DF}).}

\item{{\bf Experiment duration:} Each data point in Figs.~\ref{fig:isoTri_per_flat}-\ref{fig:isoTri_ber_flat} represents 232,000 packet transmissions, or 13.9 million packets total.}

\end{itemize}

We can make a few observations from these results. At nearly every topology point cooperation provides a performance gain; in some cases the gain is substantial. Consider the PER for QPSK payloads, shown in Fig.~\ref{fig:isoTri_per_flat}(a). The peak PER improvement for DF vs.\ NC is nearly 45 times (1.03\textsc{e}{-4} vs. 4.6\textsc{e}{-3} at -8 dB attenuation). Further, the overall shape of the cooperative curves demonstrate evidence of diversity via steeper slopes with increasing SNR than the non-cooperative curves.

It is clear that modulation rate affects the performance of every scheme, with the smaller noise margin for 16-QAM consistently degrading performance. Even subject to this performance decrease, DF (and in some cases AF) still outperform the non-cooperative link.

There are a few apparent anomalies in these results. For example, consider the PER curves for 16-QAM in Fig.~\ref{fig:isoTri_per_flat}(b). Note how the PER for AF is actually worse than NC at high SNR. But compare this to the corresponding BER curves in Fig.~\ref{fig:isoTri_ber_flat}(b). Here, AF clearly outperforms NC at all SNRs. This apparent inconsistency has a satisfying explanation and is explored in detail in Sec~\ref{sec:berDist}. Notice also how DF develops an error floor at the right-most points (i.e. at higher SNRs). This observation is explored below in Section~\ref{sec:dfCFOerrors}.

\subsection{Performance Analysis}
\label{sec:resultsDiscussion}
Overall the results of our experiments are very encouraging. In every experiment, there are points where physical layer cooperation provides significant performance gains. The overall performance of the transceiver varies as expected with channel statistics, average SNR and modulation rate.

However, a few aspects of the overall performance results presented above merit further investigation. Specifically, we seek to understand the underlying causes for performance limitations observed in our PER and BER results. These limitations manifest as error floors, regimes where performance no longer improves with increasing SNR. This section presents discussions and additional experiments exploring underlying causes of error floors in our results.

\subsubsection{Bit Error Densities}
\label{sec:berDist}
Consider the PER plots for the co-located source/relay topology in Fig.~\ref{fig:isoTri_per_flat}. In general, the DF and AF curves show significant PER improvement over NC at nearly every point. The one deviation from this general observation is for PER of 16-QAM payloads (Fig.~\ref{fig:isoTri_per_flat}(b)). Notice that at the two highest SNRs (furthest to the right) the AF curve shows worse packet error rates than NC. Compare these points to the corresponding BER values (Fig.~\ref{fig:isoTri_ber_flat}(b)). Here, both cooperative schemes significantly outperform non-cooperative. This disparity between PER and BER requires deeper investigation. 

We start by analyzing curves corresponding to each kind of packet error. Recall from Section~\ref{sec:metrics} that for every transmission, the OFDM receiver terminates in one of four states: \textsf{No Reception}, \textsf{Bad Header}, \textsf{Bad Payload} or \textsf{Good Payload}. The first three count as packet errors and our experiment design records the occurrence of each separately.

\begin{figure}[h!]
\centering
%\subfigure[PER]{\includegraphics[width=2.0in]{figs/isoTri_per_mod2_chan1.pdf}}
\subfigure[\textsf{No Reception}]{\includegraphics[width=2.0in]{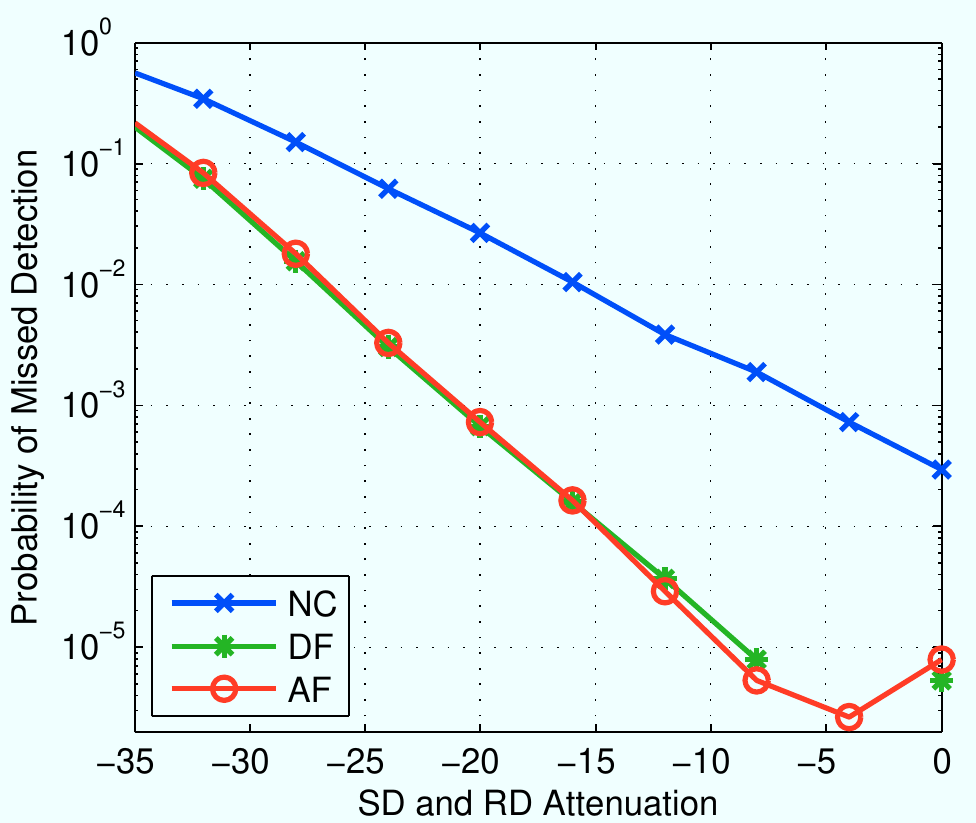}}
\subfigure[\textsf{Bad Header}]{\includegraphics[width=2.0in]{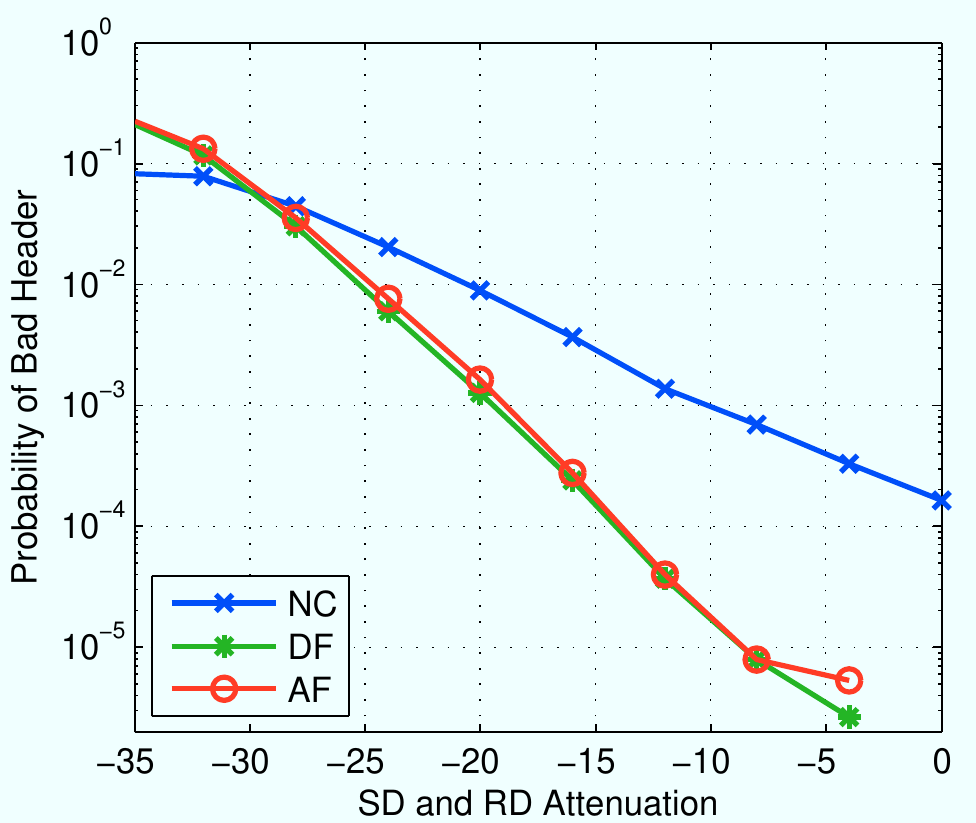}}
\subfigure[\textsf{Bad Payload}]{\includegraphics[width=2.0in]{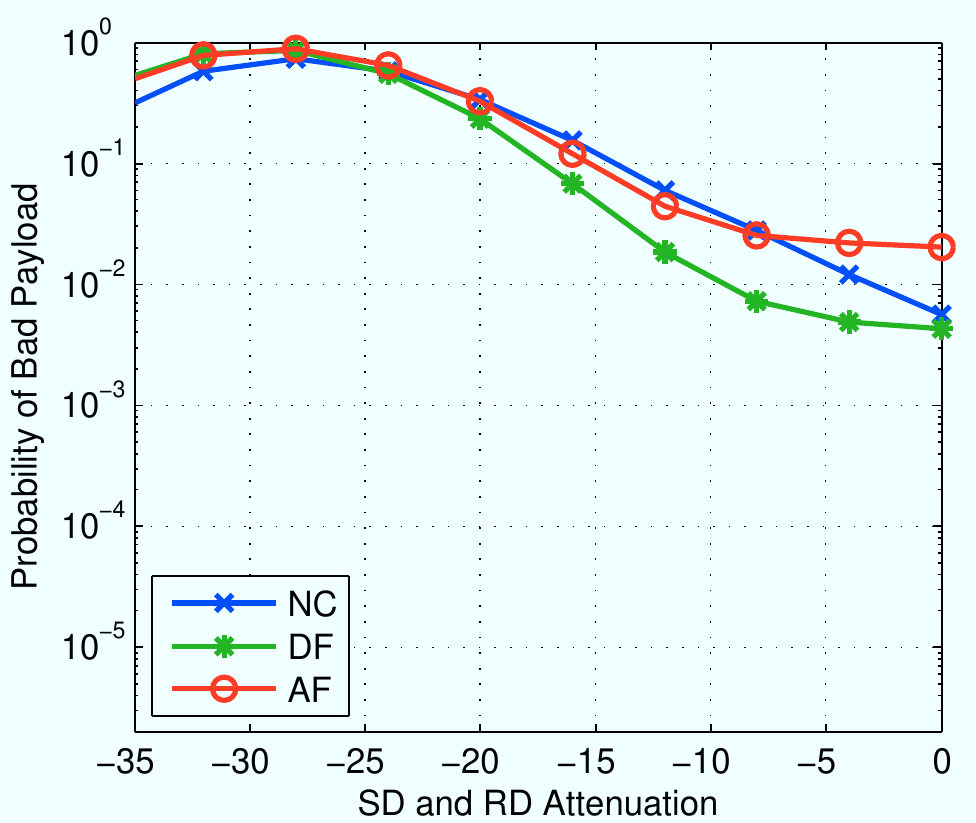}}
\caption{Probabilities of each packet error event for 16-QAM payloads in the co-located source/relay topology.}
\label{fig:isoTri_per_breakdown}
\end{figure}

Fig.~\ref{fig:isoTri_per_breakdown} presents three curves, showing the individual probabilities of the different packet error events. The sum of these three plots gives the overall PER curve in Fig.~\ref{fig:isoTri_per_flat}(b). Analyzing the individual error probabilities allows us to determine which error dominates the overall PER in various SNR regimes. At high SNR, where the PER/BER disparity for AF manifests, it is clear that payload bit errors (\textsf{Bad Payload} end states) dominate the overall PER. In other words, payload bit errors, not failed energy detection or header errors, are the dominant error source for AF in this regime. This observation leads us to dig deeper into the distribution of bit errors at the highest SNRs.

\begin{figure}[h]
\centering
\includegraphics[width=3in]{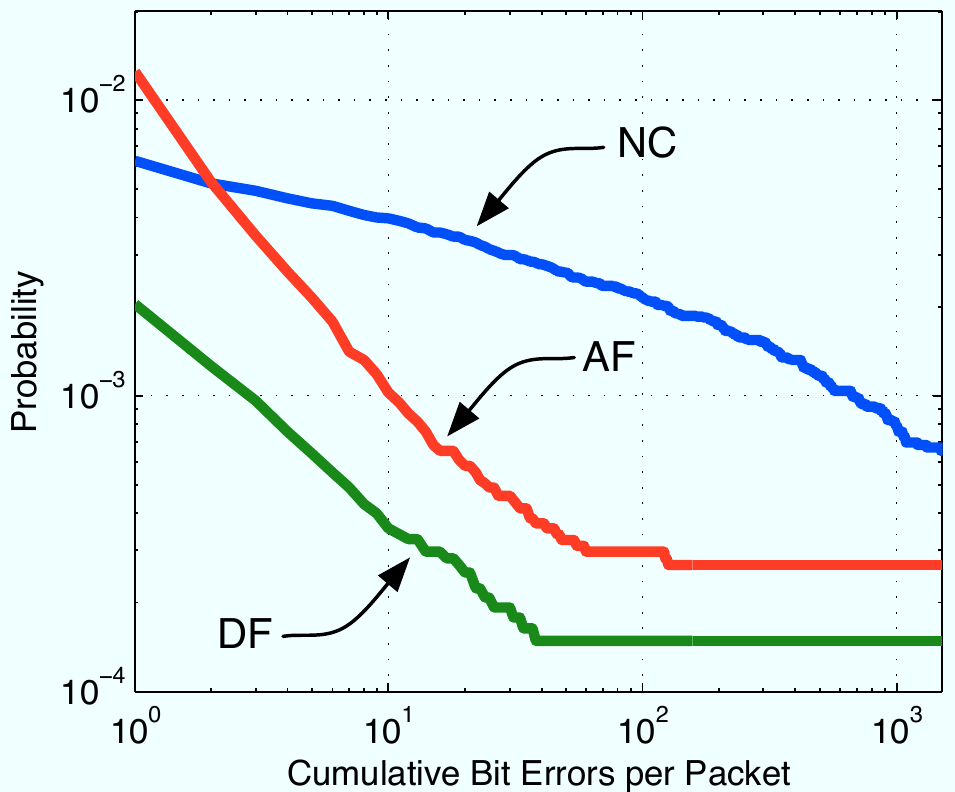}
\caption{Cumulative probability of number of bit errors per packet for highest SNR point in co-located source/relay topology with 16-QAM modulated payloads.}
\label{fig:bitErrDesnsityDist}
\end{figure}

To better understand the distribution of bit errors among packets, we designed an experiment to record the number of bit errors per packet, in addition to the overall BER. The results from this experiment, shown in Fig.~\ref{fig:bitErrDesnsityDist}, provide the data necessary to understand the PER/BER disparity discussed above. This experiment tests the co-located source/relay topology at the highest SD/RD SNR (i.e. same parameters as the right-most point in Figs.~\ref{fig:isoTri_per_flat}(b) and \ref{fig:isoTri_ber_flat}(b)). The bit error densities are plotted here as a cumulative probability, with numbers of bit errors per packet accumulating along the X-axis and probability on the Y-axis (note the Y-axis is a log scale).

Cooperation is expected to provide a diversity gain, whose key benefit is delivering packets which would otherwise be degraded or lost due to a deep fade in the SD channel. Intuitively, we should expect a non-cooperative link to have a higher probability of many bit errors per packet, corresponding to deep fades along the single fading channel. Likewise, we should expect cooperative links to deliver fewer packets with many bit errors, as a packet should only be severely degraded when two independent channels (SD and RD) are in deep fades. Both of these expectations are met by the data in Fig.~\ref{fig:bitErrDesnsityDist}. Consider the points at the right side of the plot, corresponding to the probabilities of each scheme delivering a packet with more than 1000 bit errors. The cooperative schemes clearly deliver far fewer packets with thousands of errors.

This data also helps reconcile the apparent PER/BER disparity with AF at high SNR. Consider the points at the left edge of Fig.~\ref{fig:bitErrDesnsityDist}, corresponding to the probabilities of each scheme delivering a packet with 1 or more bit error. At this one point, the AF curve exceeds that of NC. But for 2, 3 or 4 bit errors per packet (the next few points), the AF curve falls well below that for NC. This behavior is a clear demonstration of noise amplification in amplify-and-forward relays. The AF relay provides diversity, filling in deep fades and lowering the probability of many bit errors per packet. But it does so with ``noisy" power, inducing the occasional bit error where the NC link would otherwise have none. A DF relay does not manifest this behavior, as it provides ``clean" power by digitally re-generating its transmitted waveform.

\subsubsection{CFO Pre-correction Errors}
\label{sec:dfCFOerrors}

The second performance bottleneck we explore is the error floor in the DF curves in Figs.~\ref{fig:isoTri_per_flat} and~\ref{fig:isoTri_ber_flat}. This floor develops only at the highest SNRs, where the source, relay and destination are essentially co-located. To understand this floor, we first consider what operations are unique to DF in our design. Recall from Section~\ref{sec:design} that our relay implementation uses a CFO pre-correction system, and that this system is only required for DF operation. The CFO pre-correction scheme uses a CFO estimate calculated in the first time slot to pre-correct the frequency of the relay's transmission in the second slot. If the CFO estimate were in error by a large enough amount, the relay's transmission could cause a packet error, as the destination's receiver would be unable to resolve the dual carrier frequency offsets in the incoming waveform.

Recall the inter-transmitter CFO tolerance curves in Fig.~\ref{fig:berper_DF_vs_SRcfo_noPrespin} and relay CFO estimation performance curves in Fig.~\ref{fig:cfoest_dist_pilot}. These curves would suggest that for a majority of SNRs the CFO estimator performs well enough to avoid inducing errors during cooperative transmissions. However, with the introduction of fading along the SR link (i.e. randomly choosing points along the X-axis in Fig.~\ref{fig:cfoest_dist_pilot}), we should expect the CFO estimator to very occasionally calculate an erroneous value. 

\begin{figure}[h]
\centering
\includegraphics[width=3in]{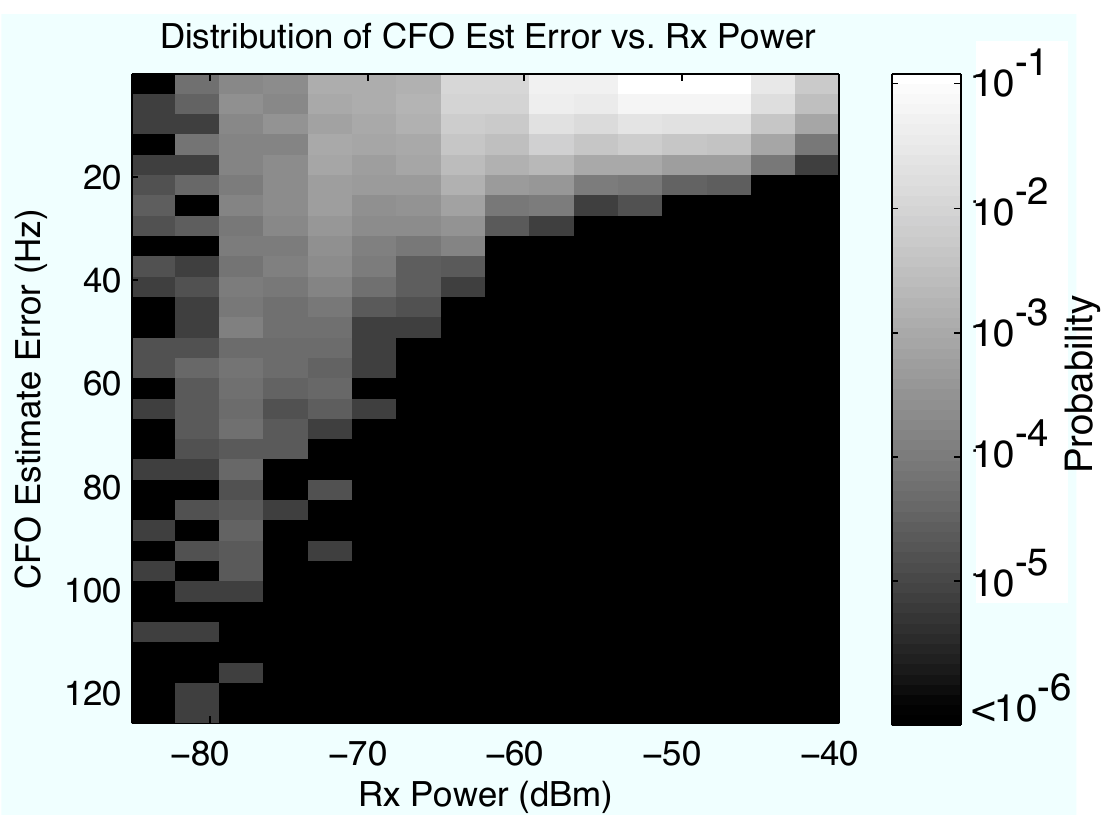}
\caption{Probability distribution of relay CFO estimation error and received power for frequency flat fading at minimum average path loss.}
\label{fig:cfo_rxpower_dist}
\end{figure}

We can test this expectation experimentally. We extended the CFO estimator characterization experiment discussed in Section~\ref{sec:cfo} to record both the CFO estimation error and received power for every packet arriving at the relay. If the hypothesis of fading causing occasional CFO estimation errors holds, we should see a strong correlation between low received power and CFO estimation error. The results of this experiment are shown in Fig.~\ref{fig:cfo_rxpower_dist}. This is a 2-D histogram showing the probability of each combination of received power (X-axis; weaker receptions to the left) and CFO estimation error (Y-axis; higher error towards the bottom). The probability of each combination is represented by the color at each point.

This plot includes only data from packets received with no errors at the relay, corresponding to those receptions which a DF relay would re-transmit. Note that the probabilities (as colors) are on a log scale. First, observe the spread of receive powers spanning 40 dB, with lower powers being less likely. This distribution represents the fading statistics of the emulator's channel model. Second, note the range of CFO estimation errors, and compare these to the PER/BER curves in Fig.~\ref{fig:berper_DF_vs_SRcfo_noPrespin}. Any estimation error is bad, but errors larger than even $\approx$50 Hz significantly degrade performance. It is clear from these results that our CFO estimator provides very good estimates for moderate-to-high SNRs. But at low SNR there is a higher probability of the estimator providing CFO values with errors large enough to degrade performance at the destination.

This observation, of rare but inevitable CFO pre-correction errors, presents an interesting dilemma. One of the attractive properties of decode-and-forward relaying is the isolation between the source-relay and relay-destination channels. In AF, for example, the re-transmitted waveform preserves whatever degradation it suffered along the SR channel. In an ideal DF relay the SR channel would have no impact on the quality of the relay's transmission to the destination; the relay strips away any received noise and channel degradations by re-generating a new, noiseless waveform for transmission.

\begin{figure}[h]
\centering
\includegraphics[width=5in]{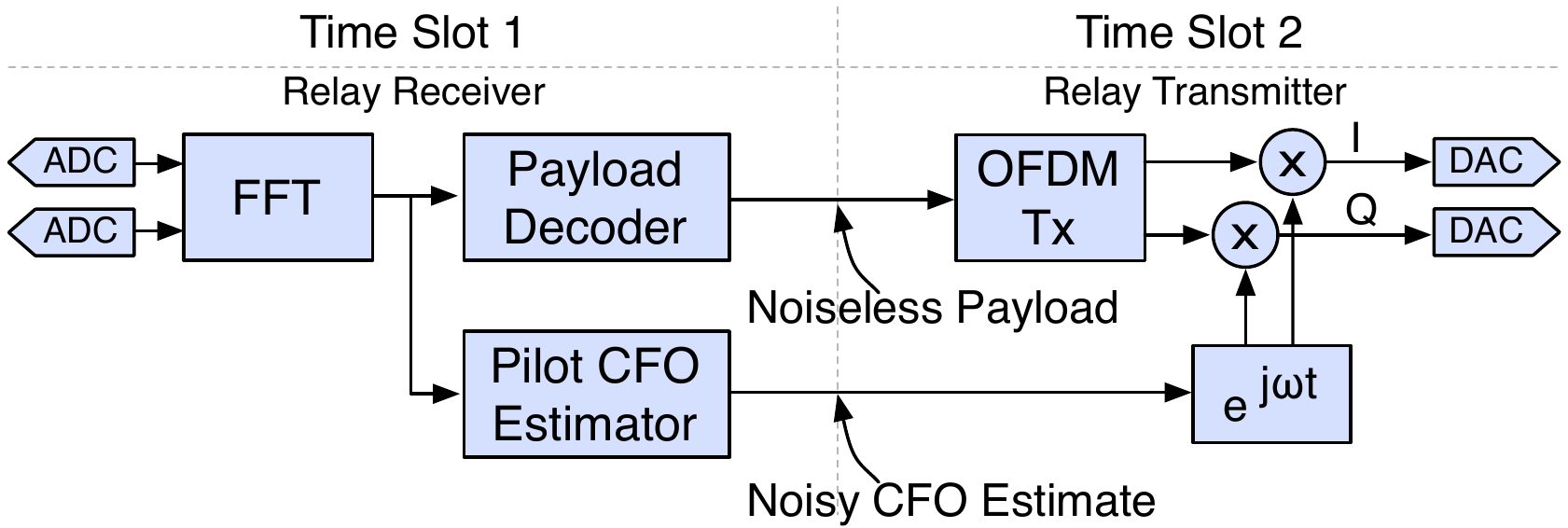}
\caption{Depiction of two things crossing the boundary between time slots in our DF relay implementation: a decoded, noiseless payload and a noisy CFO estimate.}
\label{fig:preCFO_time_blkDiagram}
\end{figure}

However, in our implementation of DF, depicted in Fig.~\ref{fig:preCFO_time_blkDiagram}, the relay applies CFO pre-correction to its transmission, and the CFO value it uses can be degraded by the SR channel. In a sense, this process allows SR channel effects to ``leak through" to the DF relay's transmission. In the worst case the relay would transmit with a bad CFO estimate and cause a packet error when none would otherwise have occurred. Thankfully, our experiments demonstrate this is a very rare event, as shown by the significant PER improvement with DF over NC in every topology. However, this effect will cause error floors at high average SNR, where packet losses due to fading are less likely and rare CFO estimation errors begin to dominate.

%%%%%%%%%%%%%%%%%%%%%%%%%%%%%%%%%%%%%%%%%%%%%%%%%%%%
\section{Experiment Results: Linear Topology}
\label{sec:linear_results}
\begin{figure}[h!]
\centering
\subfigure[QPSK payload modulation]{\includegraphics[width=2.9in]{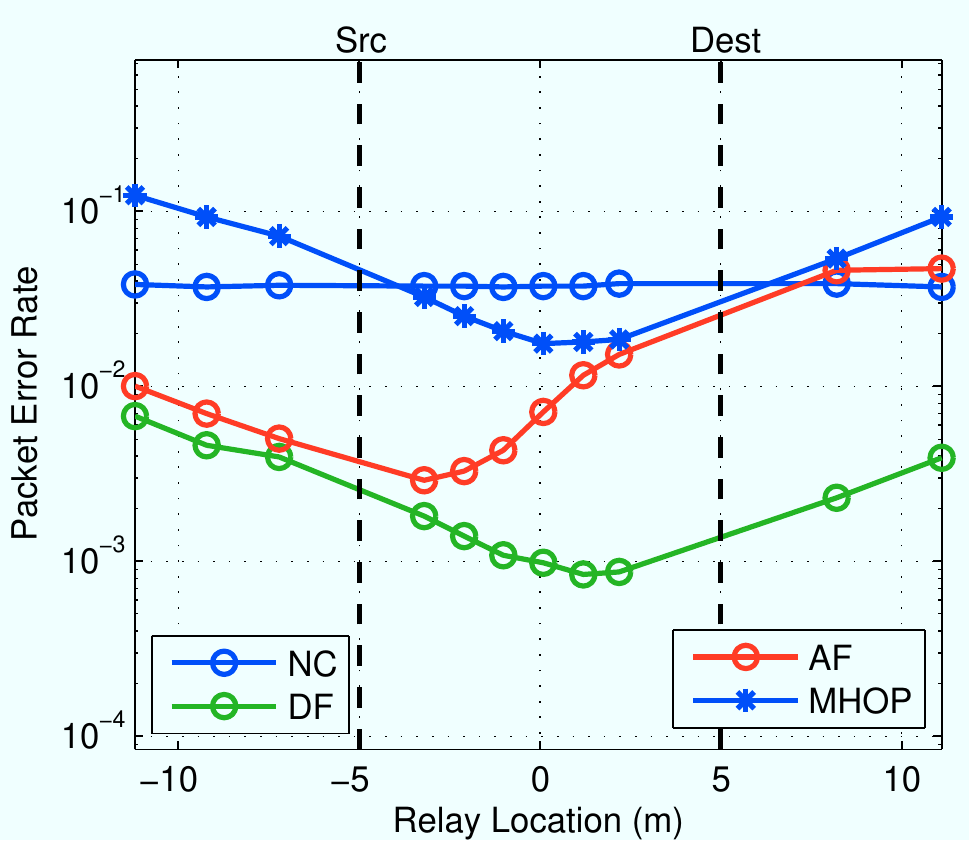}}
\subfigure[16-QAM payload modulation]{\includegraphics[width=2.9in]{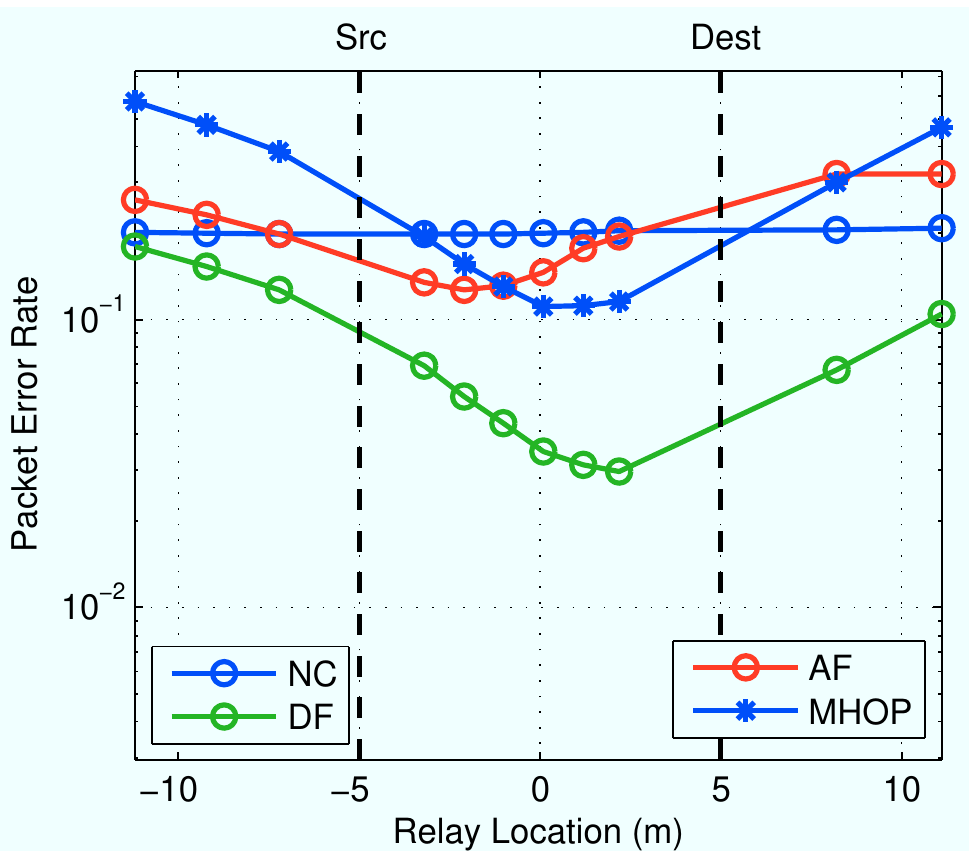}}
\caption{Packet error rates for linear topology with 1416 byte payloads in frequency flat fading.}
\label{fig:line10m_per_flat}
\end{figure}

\begin{figure}[h!]
\centering
\subfigure[QPSK payload modulation]{\includegraphics[width=2.9in]{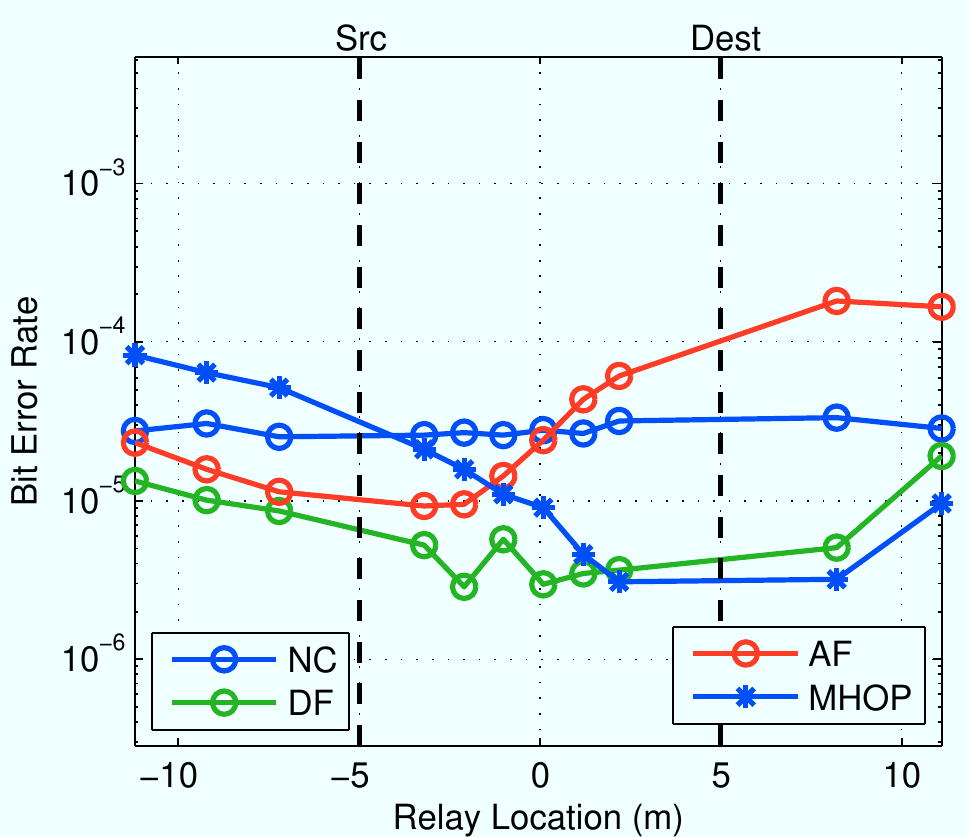}}
\subfigure[16-QAM payload modulation]{\includegraphics[width=2.9in]{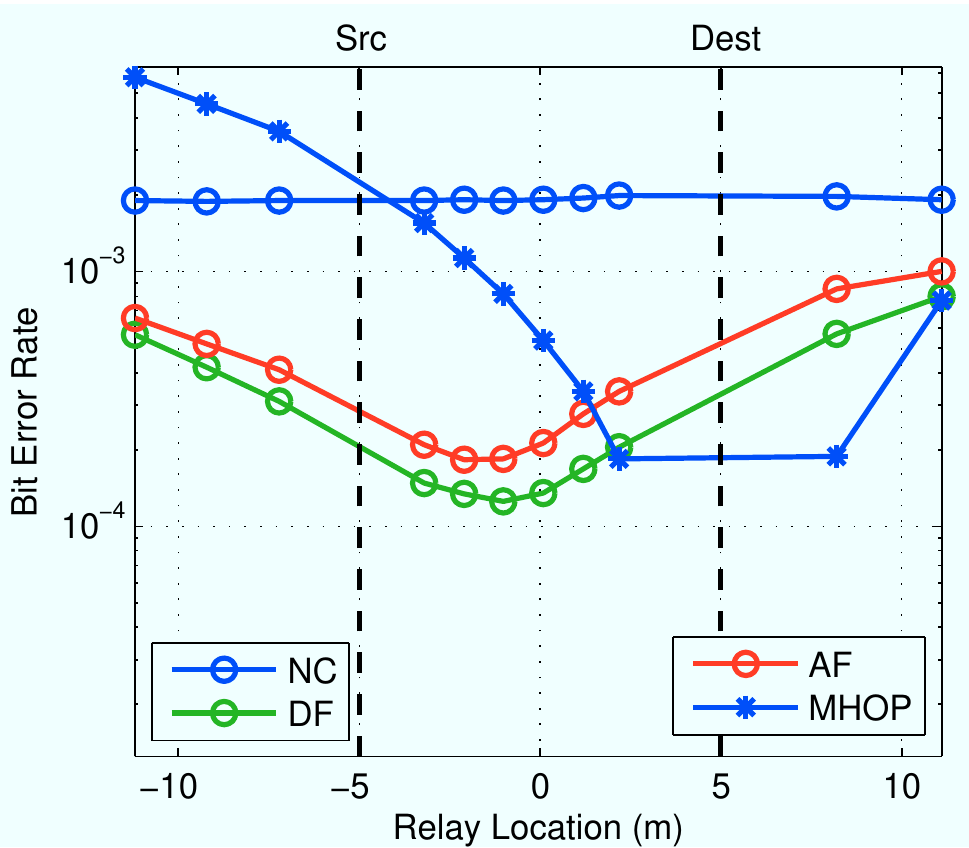}}
\caption{Bit error rates for linear topology with 1416 byte payloads in frequency flat fading.}
\label{fig:line10m_ber_flat}
\end{figure}

Finally we present results for the linear topology described above. The parameters for this experiment are:
\begin{itemize}
\item{{\bf Topology points:} This experiment tests a linear topology, emulating source-destination separations of 10.4~m. As such the SD attenuation is fixed, while the SR and RD attenuations are varied in tandem to emulate various relay positions along the SD line. In the interest of clarity the plots below use relay position as the independent variable; the actual experimental parameter is the pair of SR and RD attenuation settings.}

\item{{\bf Packet formats:} Every transmitted packet consists of a random 1412 byte payload, modulated at either QPSK or 16-QAM. Every packet has a 24 byte QPSK header. The header and payload are both uncoded.}

\item{{\bf Cooperative schemes:} This experiment tests four schemes: Amplify-and-forward ({\bf AF}), decode-and-forward ({\bf DF}), multi-hop ({\bf MHOP}) and non-cooperative ({\bf NC}). In multi-hop, only the relay transmits to the destination, and only if it receives the packet successfully from the source.}

\item{{\bf Experiment duration:} Each data point in Figs.~\ref{fig:line10m_per_flat}-\ref{fig:line10m_ber_flat} represents 319,000 packet transmissions, or 14~million packets total.}

\end{itemize}

We can make a few observations from these results. It is clear that the PER performance of DF exceeds that of all other schemes. This holds true for every relay position, channel model and modulation rate. The peak performance improvement with DF is significant. Consider Fig.~\ref{fig:line10m_per_flat}(a), which shows PER for QPSK modulated payloads. The best PER for DF is 8.5\textsc{e}{-4} (at location 1.2~m), nearly 45$\times$ better than the corresponding PER of 3.8\textsc{e}{-2} for NC. This is a dramatic improvement. That this peak improvement occurs with the relay between the source and destination makes sense. In this regime performance of the SR and RD links is balanced, allowing the relay to deliver the maximum possible assistance.

The AF curves are somewhat less impressive than DF, but still consistent with expectations. The linear topology highlights an important difference between AF and DF. In both our AF and DF implementations the relay only transmits after it successfully receives a packet from the source. While the relaying modes share this requirement, the amount of help each provides during a cooperative transmission is different. For DF, the quality of the relay transmission is determined by the accuracy of its CFO estimate. As shown in Section~\ref{sec:cfo}, the performance of our estimator is very good at \mbox{mid-to-high} SNR, and degrades seriously only at very low SNR. For AF, the quality of the relay transmission is determined by the instantaneous source-relay channel, where weaker channels result in noisier relay transmissions. 

Given that our conditions for AF and DF relay participation are the same, we can use the relative performance of each scheme to compare the impact of noise amplification (with AF) and CFO pre-correction error (with DF) as functions of relay-source separation. It is clear from our results that as the relay moves away from the source, noise amplification in AF impacts performance much more severely than CFO estimation errors in DF. These factors also help explain the relative relay locations of maximum cooperative gain with AF and DF. Performance with an AF relay peaks nearer the source. At locations beyond this point, the AF relay transmissions grow noisier and less frequent, delivering less gain with increasing source-relay separation.

In terms of PER, DF, and in most cases AF, outperform pure multi-hop. This is a clear demonstration of the benefits of diversity. Multi-hop provides no diversity improvement, succeeding only when two channels (SR and RD) are both able to successfully convey a packet. As expected, cooperation provides actual diversity, delivering packets when any combination of the SD and RD transmissions succeeds.

These experiments highlight the importance of interpreting BER and PER together. For example, note the unusual BER curves for multi-hop in Fig.~\ref{fig:line10m_ber_flat}(b). Viewed in isolation, the BER performance of multi-hop seems very good relative to the other schemes when the relay is near the destination. But recall from Section~\ref{sec:metrics} that only transmissions which end in the \textsf{Bad Payload} state at the destination contribute errors to the BER calculation. In other words, transmissions which are not detected or which end in the \textsf{Bad Header} state do not count towards BER. In the multi-hop scheme the destination can only receive packets from the relay, and the relay only transmits packets it receives from the source with zero errors. When the relay is far from the source it will only occasionally receive error-free packets. But if it is near the destination, it will successfully re-transmit these occasional packets with high probability. Thus, the multi-hop BER appears to be very good when the relay is near the destination, but the overall performance is actually very poor. This is clearly demonstrated by the corresponding PER curves for multi-hop (Fig.~\ref{fig:line10m_per_flat}) which consistently show poor multi-hop performance when the relay is near the destination.

%%%%%%%%%%%%%%%%%%%%%%%%%%%%%%%%%%%%%%%%%%%%%%%%%%%%%%%%%%%
%%%%%%%%%%%%%%%%%%%%%%%%%%%%%%%%%%%%%%%%%%%%%%%%%%%%%%%%%%%

\section{Conclusions}
\label{sec:conclusion}

We have presented the design of a complete, real-time, wideband, cooperative OFDM transceiver and extensive results detailing its performance under a variety of propagation and topological conditions. Further, we identified, isolated and explained the underlying causes of two floors observed at high SNR in our performance measurements. Our transceiver design and experimental frameworks are open-source and ready for use by other researchers, especially those exploring how to best exploit physical layer cooperation at higher network layers.

\section*{Acknowledgments}
The authors would like to thank Dr. Chris Dick at Xilinx and the Xilinx University Program for their continuing support of the WARP project, and Azimuth Systems for the use of the ACE 400WB channel emulator.

%Insert the bibliography
\bibliographystyle{IEEEtran}
\bibliography{bib/IEEEabrv,bib/murphpo}

\end{document}